# Beam Halo and Bunch Purity Monitoring

*Kay Wittenburg*
DESY, Hamburg, Germany

**Abstract**
Beam halo measurements imply measurements of beam profiles with a very high dynamic range, both in transverse and longitudinal planes. This lesson gives an overview of high dynamic range instruments for beam halo measurements. In addition, halo definitions and quantifications in view of beam instrumentation are discussed.

## 1 Introduction

This report is based on my report written for the CERN Accelerator School, Beam Diagnostics in Tuusula, 2018. I have added some more recent discussions, results and references into this report.

In particle accelerator beam experiments, background due to beam halo can mask the rare physics processes in the experiment detectors. Also, the detectors are often the most radiation sensitive components in the accelerator. The beam loss threshold enforced by the sensitivity of the experiments is often far below that imposed by the activation of machine components. The loss of < 0.1‰ particles/bunch is sometimes already critical (e.g. can be harmful). Therefore, a beam halo monitor must be capable of measuring the transverse beam profile better than this. The required dynamic range for such a monitor should be of the order of $10^5$ or even better.

Before coming to the instruments which can provide measurements with such a dynamic range, chapter 2 gives a description of the beam "halo". In the subsequent chapters transverse (Section 3) and longitudinal (Section 4) halo monitors are discussed with the focus on their high dynamic range capability. It is assumed that the reader already knows the principles of usual beam profile monitors since most of them are modified for halo measurements to get a very high dynamic range. Beam profile monitors are discussed in another lesson of this CAS.

The importance of this topic was reflected by the specialized 29th ICFA Advanced Beam Dynamics Workshop on Beam Halo Dynamics, Diagnostics, and Collimation, HALO'03, held during the week of May 19–23, 2003, at Gurney's Inn, Montauk, Long Island, New York and by the Workshop on Beam Halo Monitoring**,** September 19, 2014 at SLAC, following IBIC 2014 in Monterey.

## 2 What is the beam halo

In the summary of the HALO'03 workshop [1] is written: "…it became clear that even at this workshop (HALO'03) a general definition of "Beam Halo" could not be given, because of the very different requirements in different machines, and because of the differing perspectives of instrumentation specialists and accelerator physicists: "…from the diagnostics point of view, one thing is certainly clear – by definition halo is low density and therefore difficult to measure…". A quantification of the halo requires a more or less simultaneous measurement of the core and the halo of the beam. Halo measurements require very high dynamic range instruments and methods as well as very sensitive devices to measure the few particles in the halo.

It should be stressed that there is an important difference between beam tails and beam halo; see for example Refs. [2], [3]: Tails are deviants from the expected beam profile in the order of percent or per mille while halo goes much less than this. Figure 1 shows two examples of beam tails and beam halo to visualize this difference. Since profile measurements are often questioned at the level of a few percent, e.g. by instrumental uncertainties, the difficulty is easily seen in making halo measurements already at the level of $10^{-4}$ and beyond [4]. This lesson concentrates on halo measurements only.

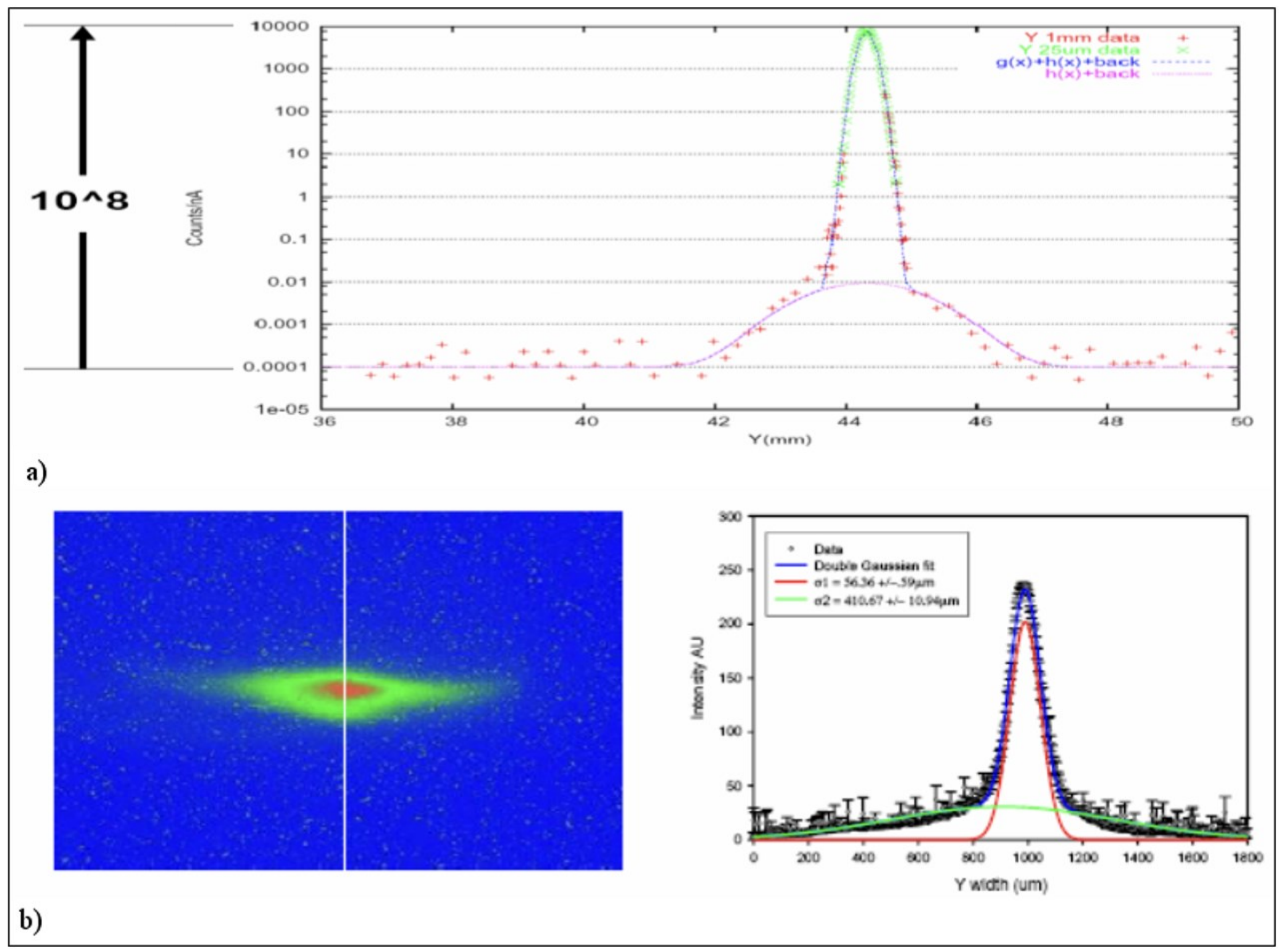

**Fig. 1:** a) Beam halo determination with a dynamic range of 108, Ref. [5], b) That's not a halo, that's a tail! Dynamic range is <103 measured with an optical transition radiation monitor [6].

### 2.1 Some sources of halo

There are numerous sources of halo formation, in linear and circular accelerators, which are summarized in [7]. Since this lesson concentrates on the halo instruments, the main sources are listed here, together with some examples of recent references in case you would like to study these effects in more detail:

- Space charge [8], [9]
- Mismatch (transversal [9] and longitudinal [10])
- Beam-Beam and nonlinear forces [11]
- RF Noise [10]
- Scattering (intra beam [8], [12], elastic and inelastic scattering, electron clouds) [13], [14]
- Diffusion [15]
- Instabilities and resonances [16].

Mostly these effects are topics of simulations which are compared with real measurements (with more or less success [17]).

### 2.2 Beam halo quantification

A measurement of the halo should result in its quantification; therefore it is important to have a definition of the halo in at least 1D spatial projections since this is typically obtained by a beam profile/halo monitor. But note that the phase-space rotations of the beam might result in oscillations of the 1D projection. For example, at some locations the halo may project strongly along the spatial coordinate and only weakly along the momentum coordinate, while at others the reverse is true, with the consequence that the halo can be hidden from the 1D spatial projections. For a complete understanding of the halo it might be necessary to extend the 1D work to the whole phase space, in the measurement (-> more locations of the monitors) as well as in the theoretical work. This leads finally to the *kinematic invariants* imposed by Hamilton's equations [18]. Such a consideration is mainly used in simulations [17], [19], [20]. This lesson concentrates on measurement issues; therefore the following concentrates on 1D projection.

In any case, the separation between the halo and the main core of the beam is not well defined. This leads to uncertainties in defining a good description of the halo content of a beam. Typically beam halo is defined as an increased population of the outer part of the beam relative to the expected distribution which describes the core (often a gaussian distribution). Five different methods for characterization of beam halo will be briefly discussed:

- Kurtosis
- q-Gaussian distribution
- Gaussian area ratio
- Ratio of beam core to offset
- Ratio of halo to core.

An important feature of such quantifiers is that they are model independent and rely only on the characteristics of the beam distribution itself. Note that a measurement always contains instrumental effects! To define the halo contents in such a theoretical way one has to exclude these effects in advance.

#### *2.2.1 Kurtosis*

This method is based on analyzing the fourth moment of the beam profile. The kurtosis is a measure of whether a data set is peaked or flat relative to a normal (Gaussian) distribution.

$$k \equiv \frac{\left\langle (x - x_0)^4 \right\rangle}{\left\langle (x - x_0)^2 \right\rangle^2} - 2 ,$$

where $x_0$ is the beam center coordinate and x is the measured value of the profile distribution. The denominator is the standard deviation of the distribution and the numerator is the fourth-order moment. The sample kurtosis of n values with mean $x_0$ is defined by:

$$k \equiv \frac{\frac{1}{n}\sum_{i=1}^{n}(x_i - x_0)^4}{\left(\frac{1}{n}\sum_{i=1}^{n}(x_i - x_0)^2\right)^2} - 2 .$$

A distribution with high kurtosis has a sharp peak near the mean that come down rapidly to heavy tails. For more details see Refs. [18], [19], [21].

### 2.2.2 *q-Gaussian distribution*

The q-Gaussian distribution is used recently to model the Halo at LHC [22]. Its probability density function is described by

$$f(u) = \frac{\sqrt{\beta}}{C_q}\left(1 - \beta(1-q)u^2\right)^{\frac{1}{1-q}},$$

with u = horizontal plane x or y and $1 < q < 3$ and $\beta > 0$. β is the distribution width and the parameter q characterizes the halo. $C_q$ is a normalization factor which depends on β, u and q. Figure 2 shows the distribution for various q and β. The normal distribution is recovered as $q \rightarrow 1$ while $q > 1$ characterizes an overpopulated halo.

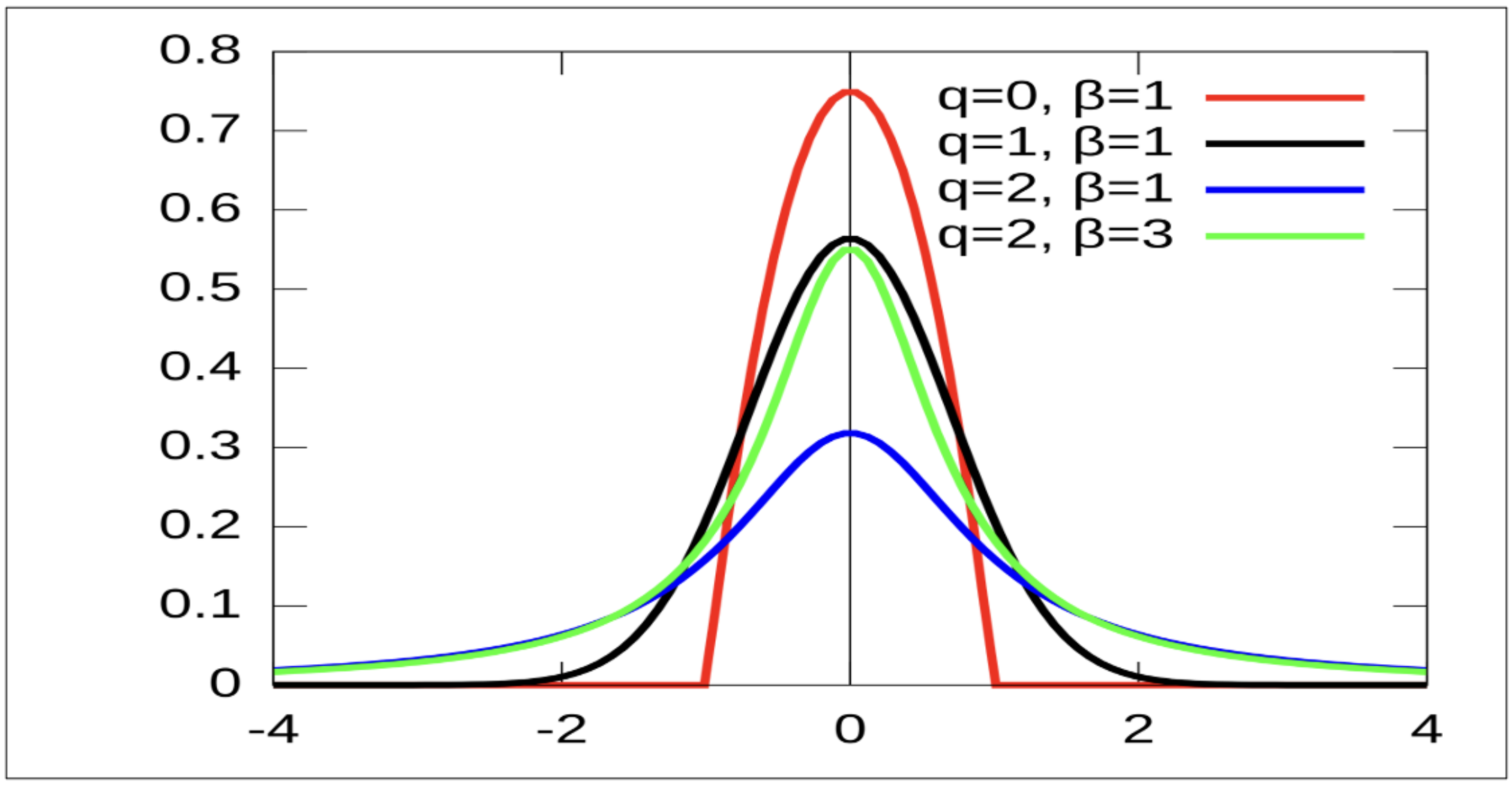

**Fig. 2:** The q-Gaussian probability density function [picture from Wikipedia: *q*-Gaussian distribution]

Both, the Kurtosis and the q-Gaussian distribution are more suited to tail than halo representation.

### 2.2.3 *Gaussian area ratio*

The Kurtosis method is quite sensitive to outlying particles, but it is not easy to apply it to experimental data. The Gaussian Area Ratio Method was found to be more useful for measurements. The method quantifies the "non-Gaussian" component of the beam profile by comparing a Gaussian fit of the core with the complete data set. Typically, the Gaussian fit is applied to the top (90 percent) of the profile to represent the core (most beam core distributions can be represented by a Gaussian). The next step is to find the integral or area of the measured distribution (e.g. by summation of the midpoint [23]) and to normalize it to (divide it by) the area under the Gaussian fit. Since the core (±1 σ) is the same in both cases one can use the area outside 1 σ only. The result (>1) gives a quantitative value of the halo content while a result =1 represents a beam without halo.

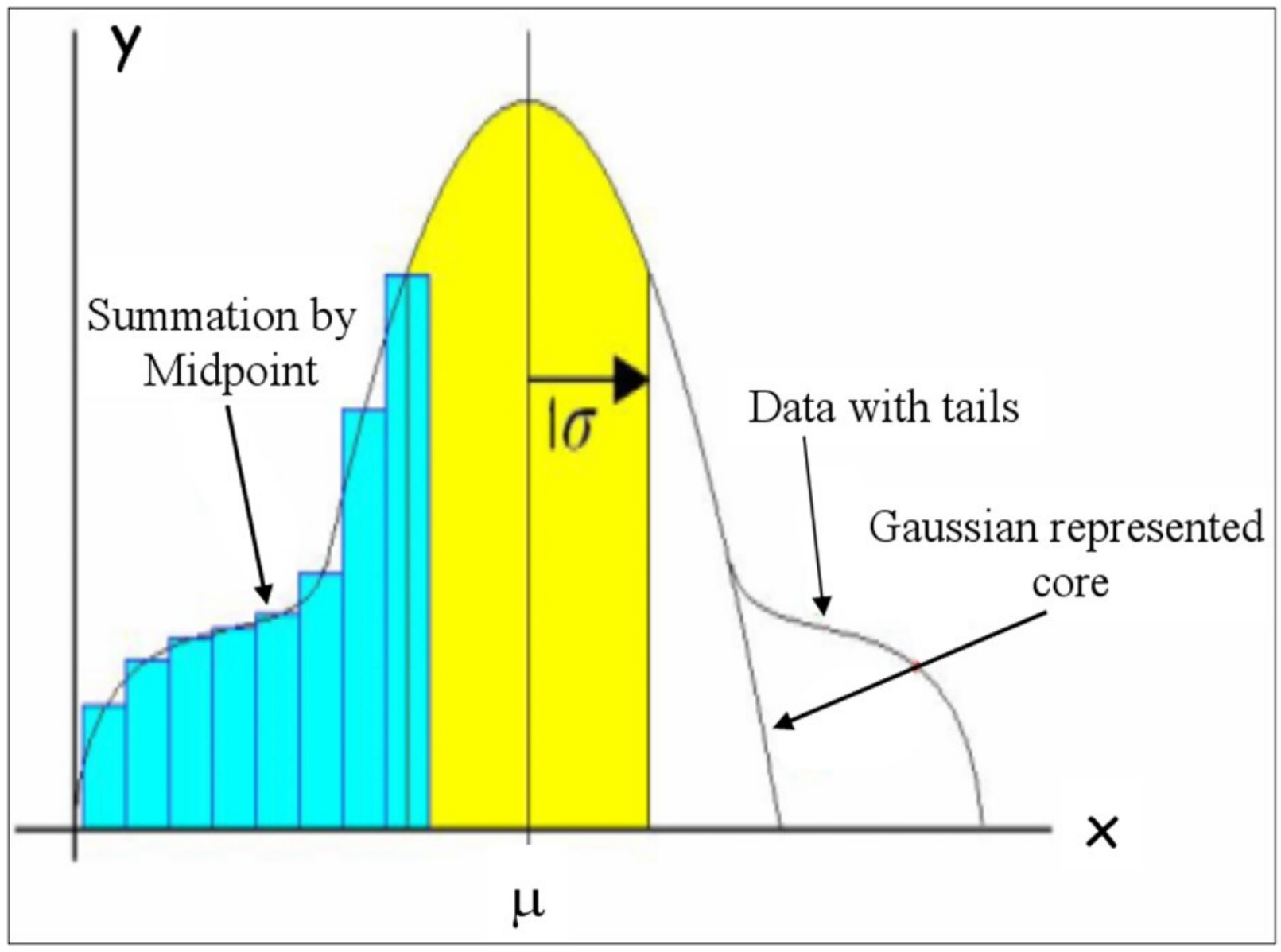

**Fig. 3:** Sketch of the Gaussian area method; from Ref. [23]

### *2.2.4 Ratio of beam core to offset*

An experimentally robust technique to quantify the halo was used at Fermilab [24]. The raw data of the detector (profile monitor) are fitted to the function:

$$f(x) = g(x) + l(x),$$

where g(x) is a Gaussian core

$$g(x) = A \cdot e^{\frac{-(x-x_0)^2}{2\sigma^2}}$$

and l(x) is the non-Gaussian halo of the beam:

$$l(x) = c_0 + c_1 x\ .$$

Defining a region of interest (ROI) which includes the tails/halo of the interesting beam profile, one can define the properties L and G as

$$L = \int_{ROI} l(x)dx$$

and

$$G = \int_{ROI} g(x)dx\ .$$

The beam halo can be calculated by the ratio L/G. A perfectly Gaussian beam will have L/G = 0, whereas a beam with halo will have L/G > 0. It is very important for this procedure to eliminate noisy and dead channels for the fit as well as knowing the pedestal for each channel. Each pedestal has to be subtracted from the data set. The standard deviation of many pedestal measurements can help to find noisy ($\sigma_{ped}$ is large) and dead channels ($\sigma_{ped}$=0). Studies [24] have shown that the L/G method is a good indicator for beam halo (see Fig. 4).

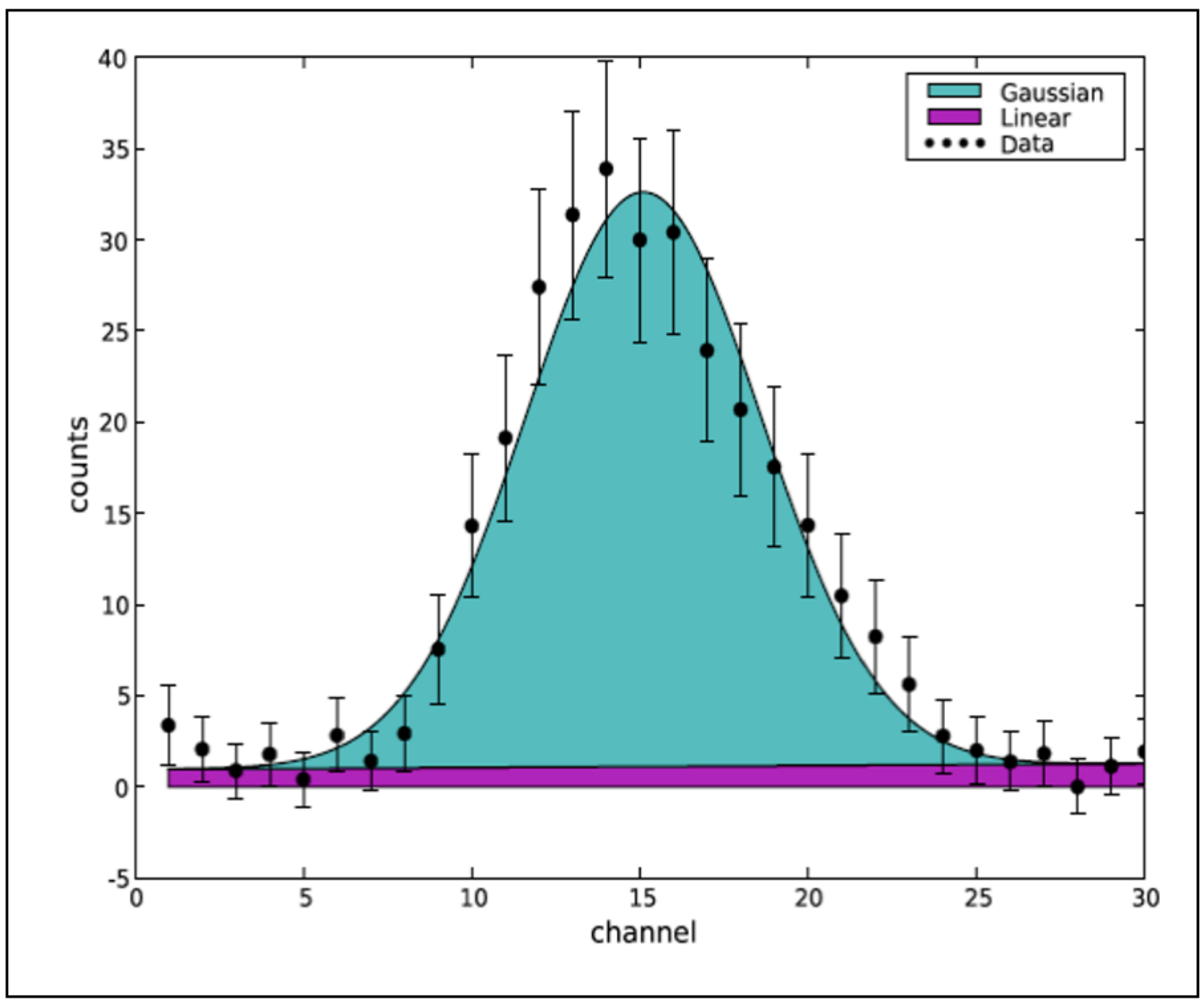

**Fig. 4:** Sketch of the Gaussian and linear fits of a beam profile; from Ref. [24].

### *2.2.5 Ratio of halo to core*

Ref [25] defined the core-halo limit as the location with the largest slope variation in the density profile. This point is well defined by the maximum of the second derivative, independent of the general shape of the distribution. This implies that the halo of a pure Gaussian profile starts at $\sigma \cdot \sqrt{3}$ and contains already 8.3% of the distribution. However, since this definition allows a clear mathematical separation between core and halo, the two quantities PHS (percentage of halo size) and PHP (percentage of halo particles) can be calculated as following:

$$PHS = 100 \cdot \frac{Halo\ sizs}{Total\ beam\ size}$$

and

$$PHP = 100 \cdot \frac{Number\ of\ particles\ in\ the\ halo}{Total\ number\ of\ particles}\ .$$

They can be calculated for any measured beam distribution along the accelerator which might change quite a lot due to non-linear effects (e.g. space charge). Neither the emittance nor the beam size (typ. 2σ) shows the development of the outer beam dimensions as clear as PHS and PHS [25]. But note that the influence of the beam tails on the definitions above might become quite strong.

## 3 Transversal halo measurements

Beam halo is the topic of different accelerator specialists:

1. Accelerator physicists: Their focus is on designing and operating their machines to minimize the halo of the accelerated and stored beam.
2. Collimation experts: These experts are interested in removing the beam halo as it appears with high efficiency and without destroying either the beam or the collimators.
3. Instrumentation specialists: Their task is twofold:

a. Providing useful information to tune an accelerator to avoid halo formation. A typical diagnostic system for this task is the tune measurement system.

b. Providing direct measurement of the beam halo. An example for a device that can directly measure halo and halo evolution is a wire scanner. An example for a device that measure more the effects of halo development is a loss monitor system.

The following lesson concentrates on the third topic, in particular on the instruments which are able to measure the beam halo and its evolution directly. Since the definition of halo is something like "$< 10^{-4}$ of the beam core", some instruments might have intrinsic limitations to get the required dynamic range. For example, ionization beam profile monitors, luminescence beam profile monitor and laser based monitors seem not yet to have sufficient dynamic range [26]. However, extensions to these devices can increase their dynamic range.

### 3.1 Halo measurements with Ionization Profile Monitor

Note the interesting idea of Ref. [27] to use an ionization beam profile monitor with additional micro channel plate (MCP) arrangement on each side of the main MCP with lower resolution but high gain for halo observations.

A recent idea from [28] is to use a masked supersonic gas jet curtain to measure the halo while masking the beam core. Both, the variation of the jet intensity and the gain of the MCP can increase the dynamic range in the halo area.

### 3.2 Halo measurements with wire scanners

Wire scanners are widely used for halo measurements with huge dynamic range and high sensitivity. This instrument provides a direct and real halo measurement by analysing the signal amplitude directly or in combination with particle counting (Section. 3.1.1). Sometimes a combination of a wire and a scraper is used to improve the sensitivity (Section 3.1.2). Typically, the signal is read out by the secondary emission (SEM) current of the wire (low beam energy) or by scintillators measuring the scattered particles (high beam energy). The problems of wire scanners are well known, e.g. emittance blow up and wire heating [29], [30].

#### *3.2.1 Wire scanner direct measurements*

The direct beam profile and halo measurement is done by correlation of the signal with the position of the wire. An example of a high dynamic range direct readout of the SEM current can be found in Ref. [31] at the PSR: A dynamic range of $10^5$ was achieved by linear amplification and the use of a 16-bit A/D converter. As an alternative solution the integrated signal was processed using a logarithmic amplifier which allows the use of a standard 12-bit A/D converter. The use of the SEM signal at low-energy accelerators has the advantage of avoiding an intrinsic error of measuring asymmetric tails by asymmetric location of external detectors and/or large beam offsets [32]. This effect vanishes at higher energies and smaller beams so that one can assume a constant solid angle during the scan. Therefore, scintillation counters outside the vacuum chamber are often used to measure the amount of scattered and shower particles created at the wire. A relatively high dynamic range can be achieved by varying the high voltage of the PMT scintillation detector [33]. Unfortunately, such scintillators are also sensitive to background due to e.g. residual gas scattering, bremsstrahlung and other sources of beam losses. A telescope counter using coincidence technique can reduce this background dramatically (see Fig. 5) as well as dark counts (noise) from the counters itself. These conditions allow the counting of the scattered particles when the wire is in the tails/halo of the beam at a constant position. Examples of this technique can be found in [34], [35]; a dynamic range of $10^8$ was achieved by [5], [34]. Under the very clean conditions of the HERA proton storage ring with very long beam lifetime (> 1000h) and therefore with nearly no beam loss a simple scintillator was sufficient. In Ref. [36] a dynamic range of

> $10^7$ was reported with halo measurements down to 6σ beam size. The lower limit is defined by the remaining background rate.

But note a few constraints of these methods:

- The counts have always to be normalized to the time interval of the measurement (wire at a constant position). Especially at very low count rates the time interval of the measurement can be very long (>10 min or so).
- The count rate has to be normalized to the beam current and beam position in case it changes during the measurement.
- Saturation effects arise if the count rate reaches 1/10th of the bunch crossing rate. This has to be taken into account. Near the bunch crossing rate (or with DC beams near limitations defined by the electronics) complete saturation occurs. Therefore, this method might be not applicable near the core of the beam. Different wire diameters can extend the useful range of this method [34].
- Beside saturation of the count rate in the beam core the wire might not survive at a constant position due to wire heating. Therefore, it should be scanned much faster through the beam and "normal" amplitude readout has to be used. In this case one needs a sufficient overlap between the two readout methods (amplitude and counting) to normalize them to each other to get the halo content relative to the beam core. A detailed analysis can be found in [36] (see also Fig. 6).

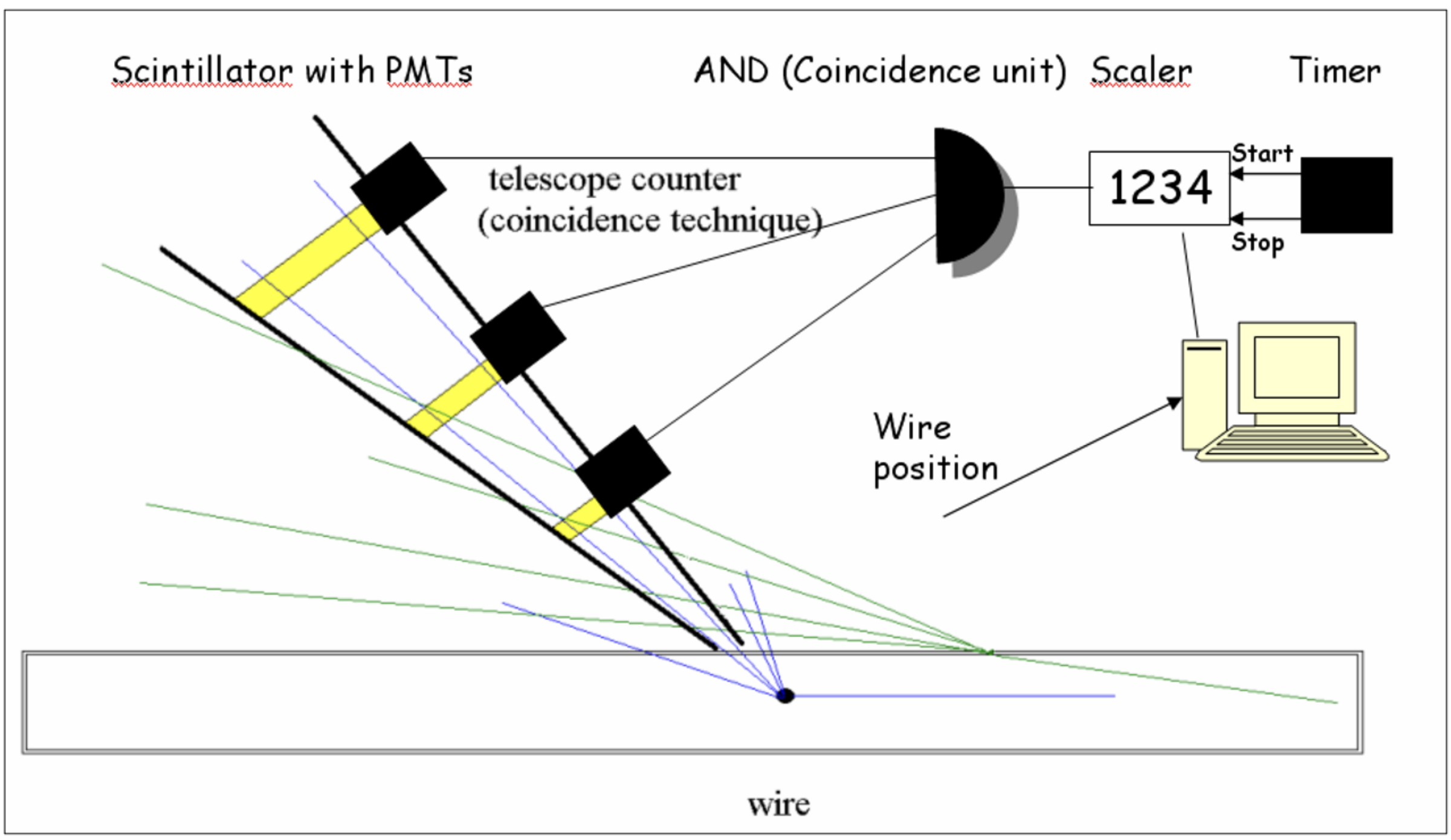

**Fig. 5:** Sketch of a telescope counter using the coincidence technique. The signals of the three scintillators are read out by photomultipliers (PMTs). Their signals are converted to logical signals and compared in a coincidence unit. A signal is counted only if all three scintillators counters give a signal, indicating that a scattered particle had crossed the three scintillators. Therefore, the telescope has to be directed to the wire. This avoids counts from other sources, e.g. beam losses.

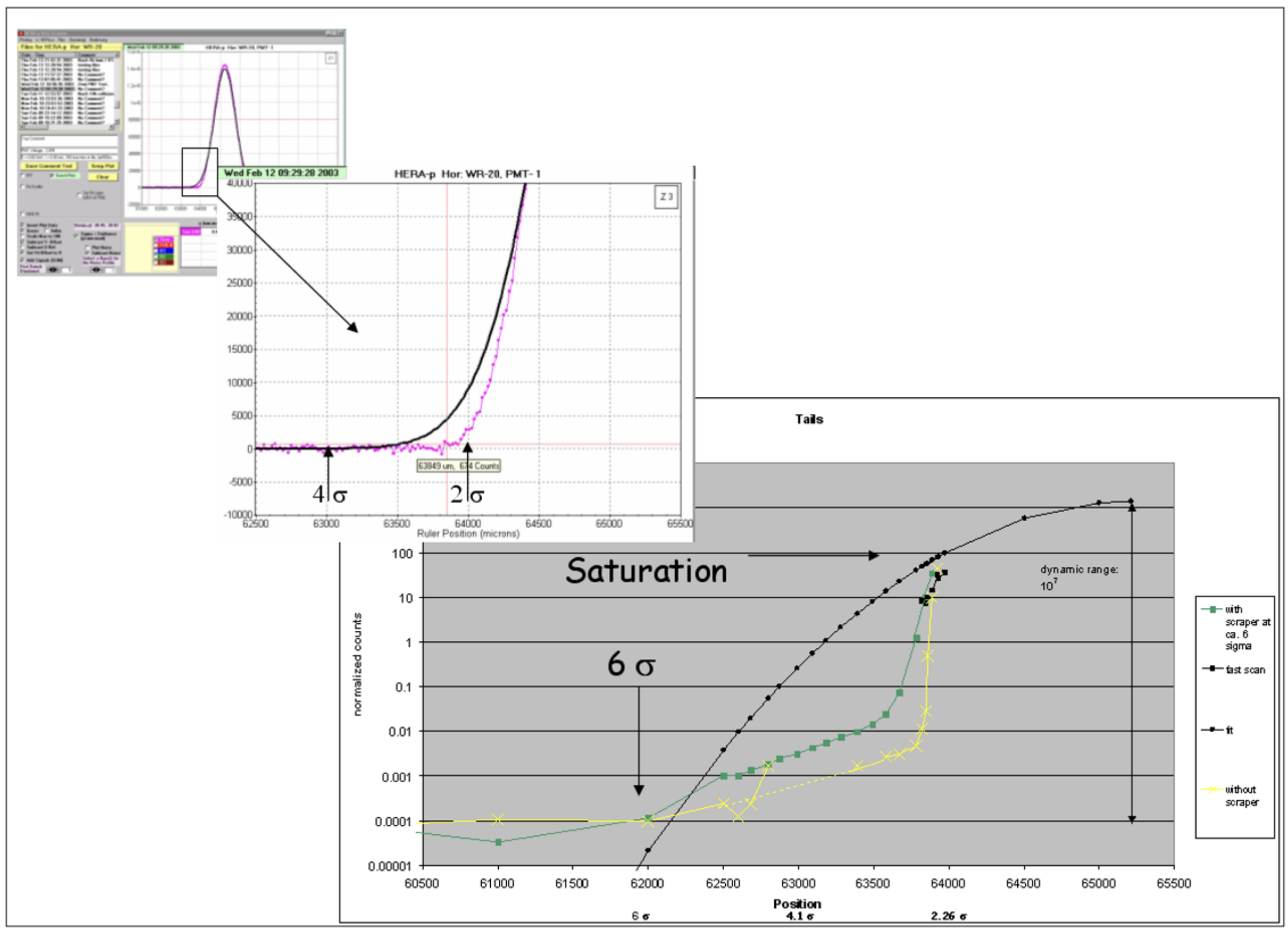

**Fig. 6:** Zoom of the tail and halo of a very stable proton beam in HERA. In the two upper figures the red (and noisy) line is the measured profile during a fast scan while the black (and smooth) line is a Gaussian fit to the data. The same fit is shown in the lower figure also in black (marked with ●) (note the log. scale). The two other lines (marked by green ■ and yellow **X**) are two halo measurements by counting signals. Indicated by → is the saturation rate and by ↓ is the lower limit background rate at 6 σ. Near the saturation at about 2.3 σ are black ■ marks indicating the (small) overlapping area of both readout methods. Note that in this result the beam tails and the beam halo are even smaller than the Gaussian fit of the beam core. From Ref. [36].

### *3.2.2 Wire scanners + scraper*

At low energy accelerators and/or at low bunch repetition rates like in a LINAC the counting method might not be very useful. In addition, a SEM current read-out of a thin wire in the beam halo will not deliver enough current for a reliable measurement. Therefore, the wire size has to be increased even to a solid scraper to increase the achievable signal. Such a scanner/scraper combination is used at LEDA [37]. The scanner consists of a 33-µm carbon fiber and two halo scrapers consisting of two graphite plates (one for each side of the distribution, see Fig. 7). Special care has to be taken that the beam does not induce too much heating of the scraper. High-heat flux testing has shown that it is possible to produce a design that can withstand the thermally induced fatigue loading even at high current operation [38]. Like in the counting method, the wire scanner and two scraper data sets must be joined to plot the complete beam distribution for each axis. The procedure is described in detail in Ref. [39]:

- Scraper data are spatially differentiated and averaged. As the scraper marches inward, it intercepts with an ever increasing segment of the beam. It is therefore necessary to differentiate the scraper signal in order to determine the transverse distribution.
- A sufficient spatial overlap is needed between the wire and scraper data. The overlap region starts with the wire positions with a sufficient signal-to-noise ratio of the wire data up to the maximum insertion location of the scraper.

- The differentiated scraper data must be normalized to attach it to the wire scanner data. The scaling factor is determined by the overlap region. The resulting three distributions are now combined into a single distribution.
- Normalize data to the position axis. Here the mechanical tolerance between the wire and the scraper position has to be taken into account.
- Normalize data to the beam current and beam position (true for all kind of halo measurements). Each data point has to be normalized to the measured beam current and beam position for each measurement.
- Before the scan: A definition of safe scraper insertion limits to avoid too much heat load on the scrapers. Thermal electron emission has to be avoided because the resulting current is orders of magnitude larger than the useable SEM current [40].

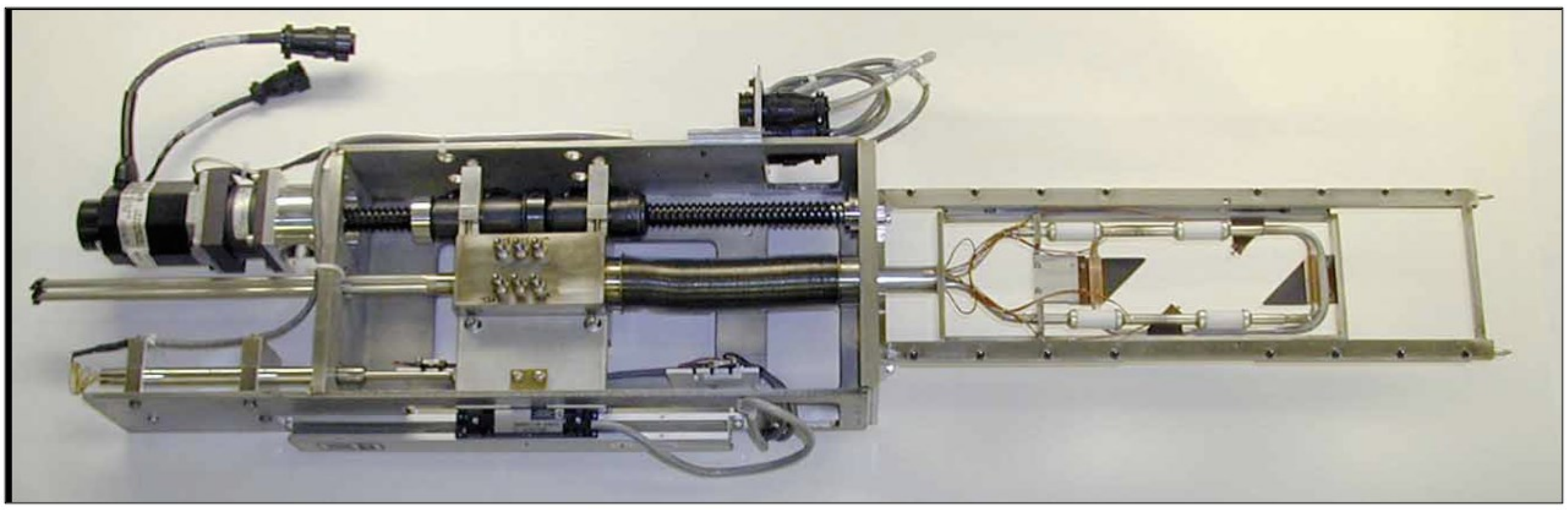

**Fig.7:** The LEDA scanner/scraper combination (from Ref. [37]).

Diamond sensors became very popular for halo measurements in the last years [41]. These devices have a huge dynamic range up to 11 orders of magnitude with photons and expected 7 orders of magnitude with electrons [42], [43]. At FACET-II [42] it is planned to extend a wire scanner with Diamond sensors for very high dynamic range halo measurements (see Fig. 8). Such an extent enables a full beam scan down to the halo region.

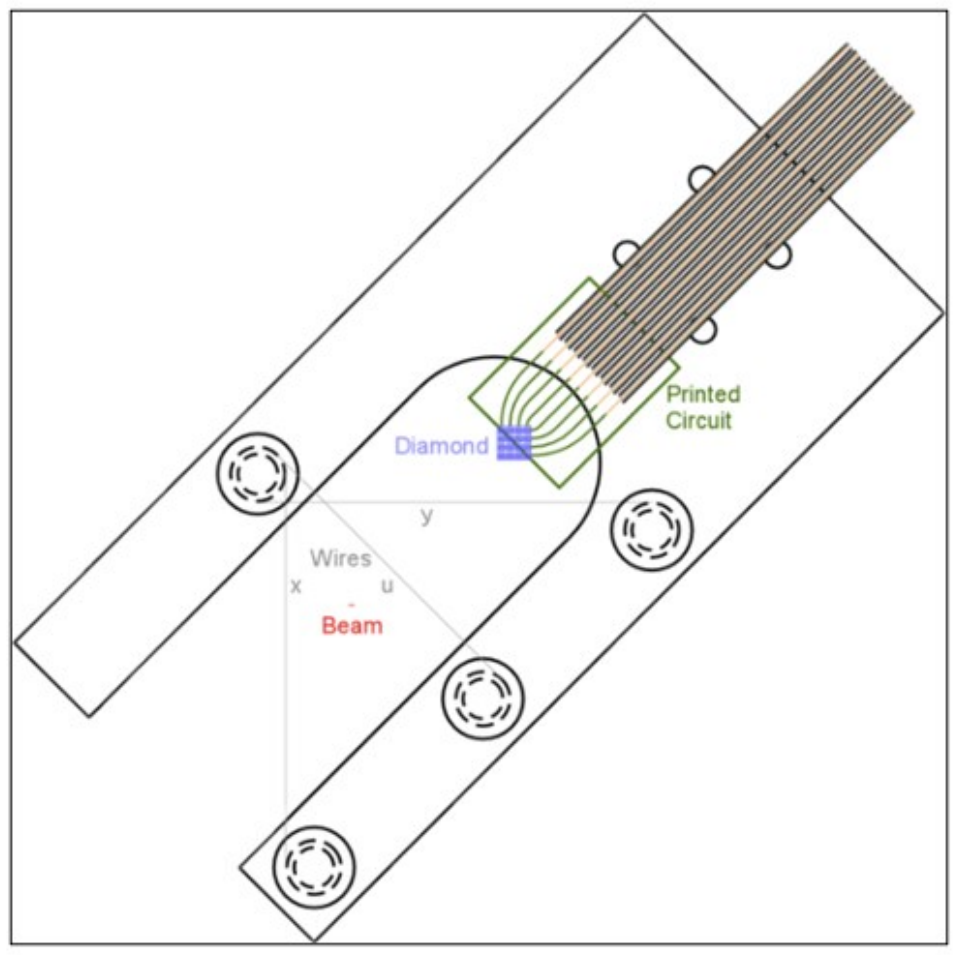

**Fig. 8:** Sketch of a modified wire-scanner harp [42]

### 3.3 Halo measurements by scraping with collimators

In a synchrotron, one jaw of a collimator will always scrape both sides of the beam distribution in one plane because of the β-oscillation of the beam particles. Therefore, one will always measure a symmetric distribution. A scan yields information about particles which oscillate with amplitudes larger than the current position of the collimator. Measurements can be performed by moving one jaw of a collimator closer to the beam in steps. Either the beam current [44] or the signal from adjacent loss monitors can be recorded for each collimator setting. The derivate of the signal will give the beam distribution. An early device using the beam current degradation while scanning the whole beam was the so-called “beam scope” [45]. Much more sensitivity can be achieved by using beam loss monitors (BLMs) close to the collimator jaws [46]. The signal of the BLMs is proportional to the inverse lifetime of the beam which gives loss rates directly in terms of equivalent lifetimes. By moving the collimators closer until significant lifetime reductions are observed, the lifetimes calculated from beam currents can be used to calibrate the BLMs. Using the BLM signals even very small beam halos created by the scattering of electrons on black body radiation were measured in LEP [47]. Other measurements using BLMs in the beam halo are reported in the Beam Loss Monitor session of this school. Since this scraping method is also a slow process it is very important to normalize each data point to the measured beam current (which changes because of scraping), to the measured beam size of the beam core (often constant) and to the beam position! In Ref. [46] a few examples are given which show artificial asymmetric tails (scanning from both directions) due to orbit movement during the scans.

Note that in high energy and/or high intensity accelerators/storage rings a complete scan of the whole beam is impossible. Collimator jaws are typically not designed to withstand the full beam. Therefore, a calibration of the halo contents (relative to the beam core) is often not possible or contains large errors. But relative changes of the halo can be detected at a very low level and far outside the beam core.

### 3.4 Miscellaneous, slow scanning or intercepting

There are other devices based on the scraping or intercepting mechanism to observe the beam halo which will not be discussed here in detail. Older and recent examples can be found in Refs. [48]-[50]. An enormously sensitive device for halo observation is the “vibrating wire scanner” [51], [52]. This very slow scanner uses as the signal source the wire position versus the change of the resonance frequency of a stretched wire due to its heating by the beam. Since a frequency change can be detected very precisely, it can be used to measure the halo directly with a very good resolution. These devices all have the same strong limitation in determining the halo relative to the beam core, but relative changes of the halo can be observed with high sensitivity and resolution.

### 3.5 Optical methods

The previously discussed methods for measuring the halo distribution are relatively slow. Scanning of the halo typically needs seconds to minutes. One needs a stable beam, and precise correlations with the beam size and position are mandatory. Optical methods, e.g. light generation by synchrotron radiation, optical transition radiation, etc. can give enough light to measure the core of even one single bunch at one passage. The light generation of these effects is linear over a huge dynamic range; the dynamic range is limited only by the light detector. In the following, four different optical methods are discussed which have the potential to improve the speed limitation:

- Two simultaneously sampling CCDs
- CID camera system
- Micro Mirror Array
- Halo measurements with coronagraph.

Using simple scintillator screens might suffer from saturation effects of the light detector and of the screen itself. However, these screens are sensitive enough to provide good signals in the beam halo [53]. To overcome the saturation effects, different screen materials with different sensitivities can be inserted into the beam while the least sensitive material (OTR) is used in the beam core, more sensitive materials in the tails and maximum sensitive material (scintillator) in the beam halo. Recent results with such a setup are reported in Refs. [54] and [55].

### *3.5.1 Two simultaneous sampling CCDs*

Commercial 8-bit CCD cameras are typically sufficient for beam profile monitors. Obviously, their dynamic range is limited to about $2{\cdot}10^2$. The idea of Ref. [56] was to use a camera with beam splitter and two CCD sensors well aligned to each other [57]. The camera has independent electronic gain control and integration time control for both sensors. The gains and integration times were set in a way that some fraction of the second CCD sensor will be saturated. Charge bleeding to the neighboring pixels has taken into account but not observed. Care has been taken to produce images on the two sensors with enough overlap to combine the data into one high dynamic range measurement. Figure 9 shows the result of such a combination with a dynamic range of $5{\cdot}10^4$. The combining algorithm is efficient enough to provide 5 Hz repetition rate for 1024x768 images.

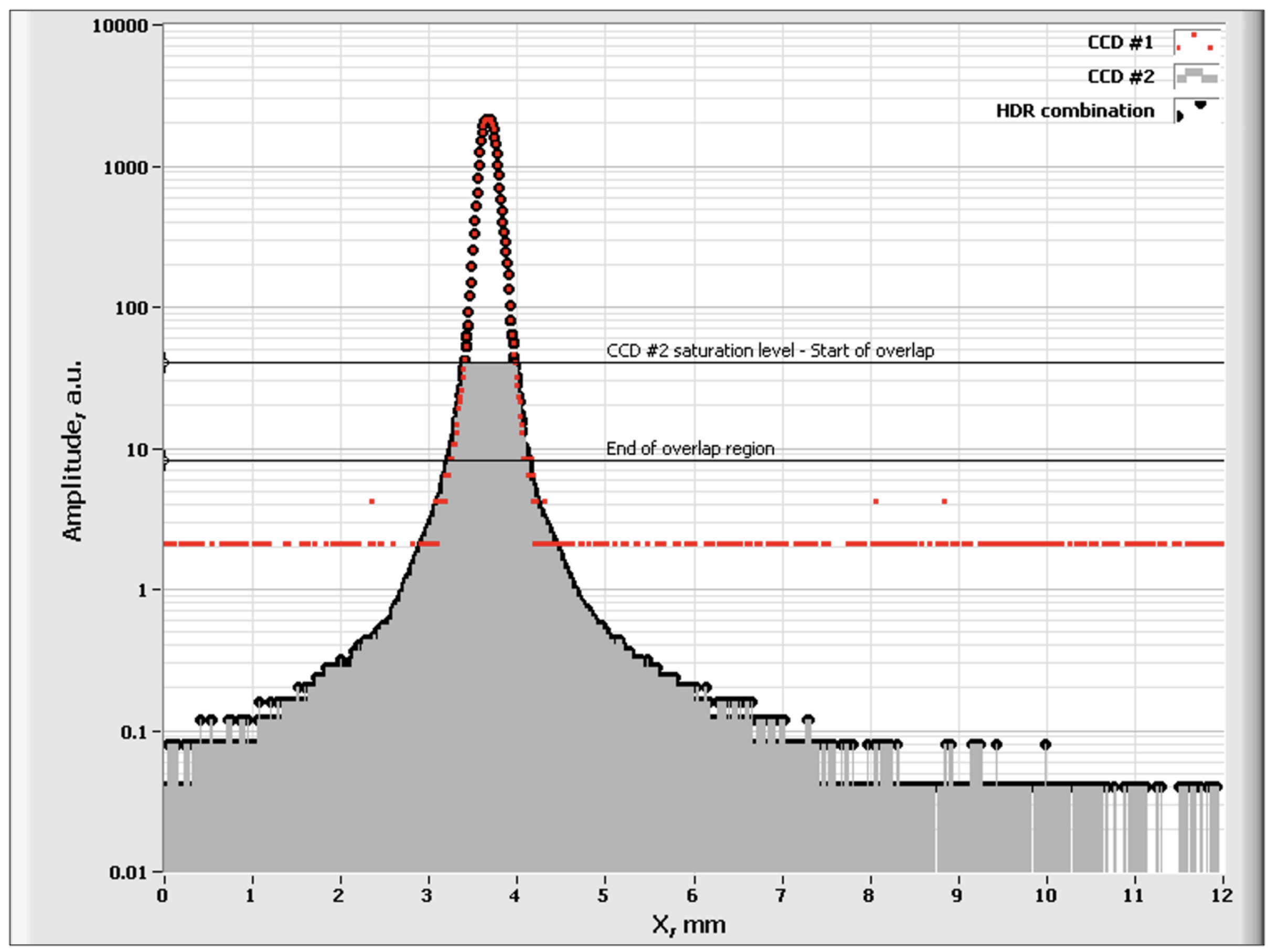

**Fig. 9:** HDR algorithm example, from [56].

### *3.5.2 CID camera system*

A camera system based on charge injection device (CID) technology can extend the dynamic range to $10^8$. The CID array consists of light sensitive pixel sensors, like a CCD array. But each pixel on a CID array is individually addressable and allows non-destructive pixel readout. In the *fixed time mode* of a CID camera the integration time of predefined regions can be adjusted based upon the expected photon flux. Therefore, different regions of the array can have different integration times, e.g. the core of the distribution should have a much shorter integration time than the tails and very much shorter than the halo. By folding the known integration time into the measured signal, the dynamic range of the whole image can be extended by a few orders of magnitude.

The *random access integration* (RAI) mode automatically adjusts the integration time based upon the photon intensity. One can predefine a most illuminated region (e.g. the core) and one or more larger regions of interest (ROI). Whenever the signal in the most illuminated region reaches a predefined threshold signal level (typically 75% of full capacity), the ROIs are read out automatically and each pixel data is added individually in the camera memory. Then the pixels in the arrays are reset and a new integration cycle begins. In this way the cycle time is automatically adjusted based on the signal accumulation rate in the most illuminated region. Each ROI can be cycled as many times as necessary until a user defined exposure time expires or until the camera memory is full. This RAI mode allows a dynamic range of $>10^6$. But note that the cycle time may last even minutes to generate enough accumulated signals in the halo region. Therefore the same requirements on beam stability are valid than for scanning devices. But a real simultaneous measurement of the core and the halo is provided.

Such a CID camera system is commercially available [58]. More details and operational experience with a CID camera system are reported in Refs. [59]-[62].

### *3.5.3 Micro mirror array*

A method to extend the dynamic range of a CCD detector system, for example, is to make it as sensitive as possible and to make sure that at this condition the core of the distribution will not arrive at the detector. The core can be measured in a second stage with much lower (but known) gain, e.g. with the help of precise optical filters. Then the relative amplitude between halo and core can be worked out easily. Since the beam may change its position and size under different conditions, one may run into problems with a fixed mask in the optical system. An adjustable Micro Mirror Array (MMA) allows the adjustment of each of the up to 11-megapixel micro-mirrors individually. Each mirror in the array is 8µm x 8µm and can be individually tilted by the high-speed integrated CMOS circuitry underneath the array, see Fig. 10 and Ref. [63]. They can be rotated by $\pm 10^0$ within about 15 ms. The first applications were in digital projection equipment, which has now expanded into digital cinema projectors. Recently MMAs have been studied for beam halo measurements at CLIC [64], [65], at CTF3 [66], at Spear [67] and UMER [68]. The process for a halo measurement can be done in the following way: First a normal profile is sampled which defines the core and the mask size. The corresponding mirrors are adjusted to tilt the core of the distribution out of the detector. Then a re-measurement of the distribution is done with high gain to view the beam halo. Therefore at least 2 successive measurements with stable beam conditions are needed. Variations of the beam in size and position can be considered by readjusting the MMA.

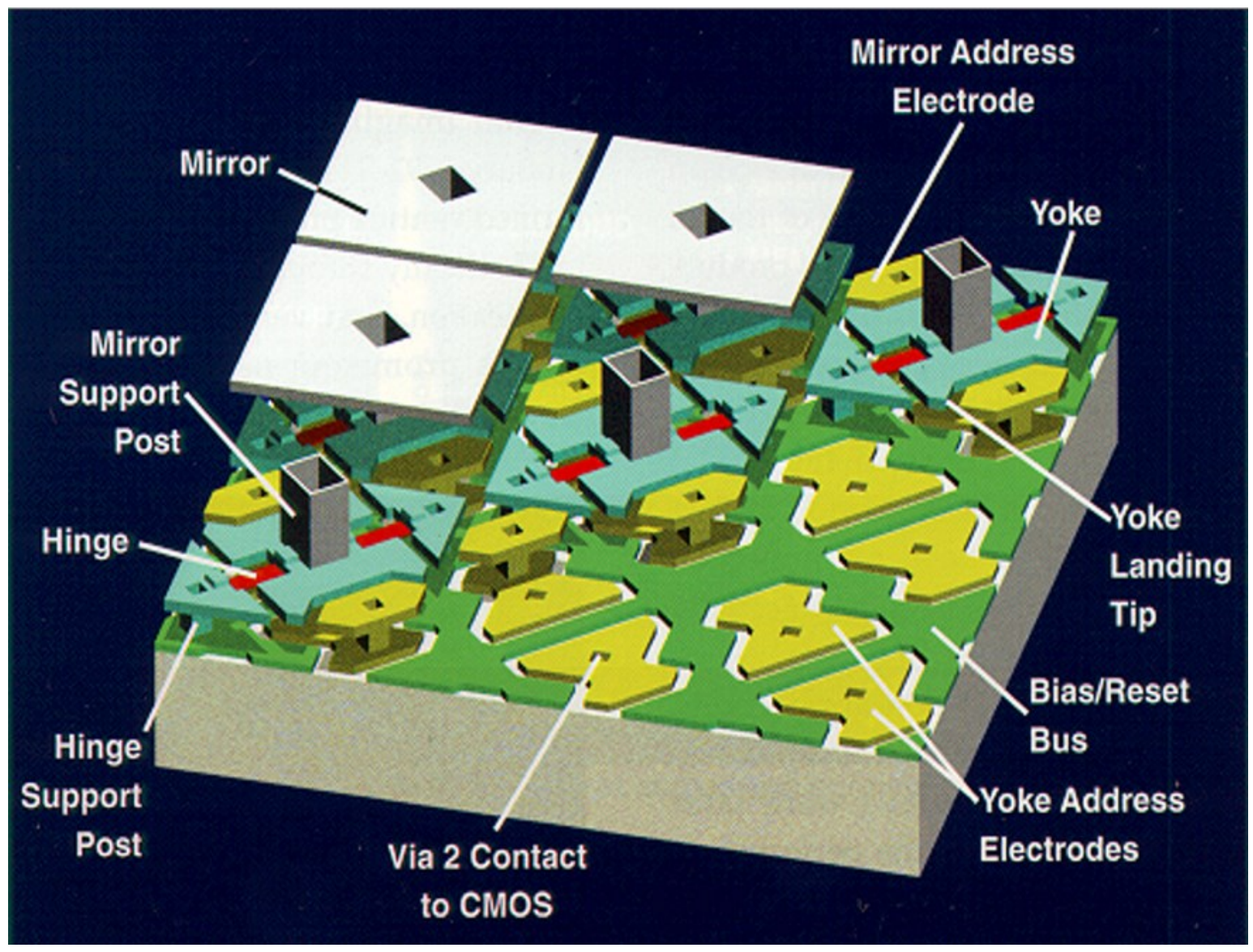

**Fig. 10**: Micro Mirror array (from Ref. [64]).

### *3.5.4 Halo measurements with coronagraph*

Observing the beam with optical radiation (e.g. synchrotron radiation or optical transition radiation) with small opening angles ($\approx 1/\gamma$) suffer from diffraction limits: The diffraction fringes produce tails surrounding the central beam image. The intensity of diffraction tails is in the range of $10^{-2}$ -$10^{-3}$ of the peak intensity which makes halo observations of lower than $10^{-3}$ impossible. A simple masking of the core will not remove the fringes (see Fig. 11).

A coronagraph is a telescope designed specifically to block out the direct light from a star, to observe nearby objects without burning the telescope's optics. A coronagraph with a so called 'Lyot stop' was developed in 1930 by the astronomer Bernard Lyot [69]. The Lyot stop is a special mask to remove the diffraction fringe while the resulting image is relayed by a special optics onto the final observation plane. A detailed description of the device and its application for beam halo measurement can be found in Ref. [70]; more beam applications can be found in Refs. [71]-[74]. A background level of $6{\cdot}10^{-7}$ at an exposure time of 100 ms and a spatial resolution of 50 μm was achieved. For such a low level all sources of background had to be minimized or eliminated. Background sources were:

- Scattering by defects on the lens surface (inside) such as scratches and digs.
- Scattering from the optical components (mirrors) near to the coronagraph.
- Reflections in the inside wall of the coronagraph. -> Cover the inside wall with light trapping material.
- Scattering from dust in the air. -> Use the coronagraph in a clean room.

Other limitations are:

- The light intensity has to be intense enough to explore the halo distribution.
- The beam position and size have to be stable to avoid saturation of the detector (camera).

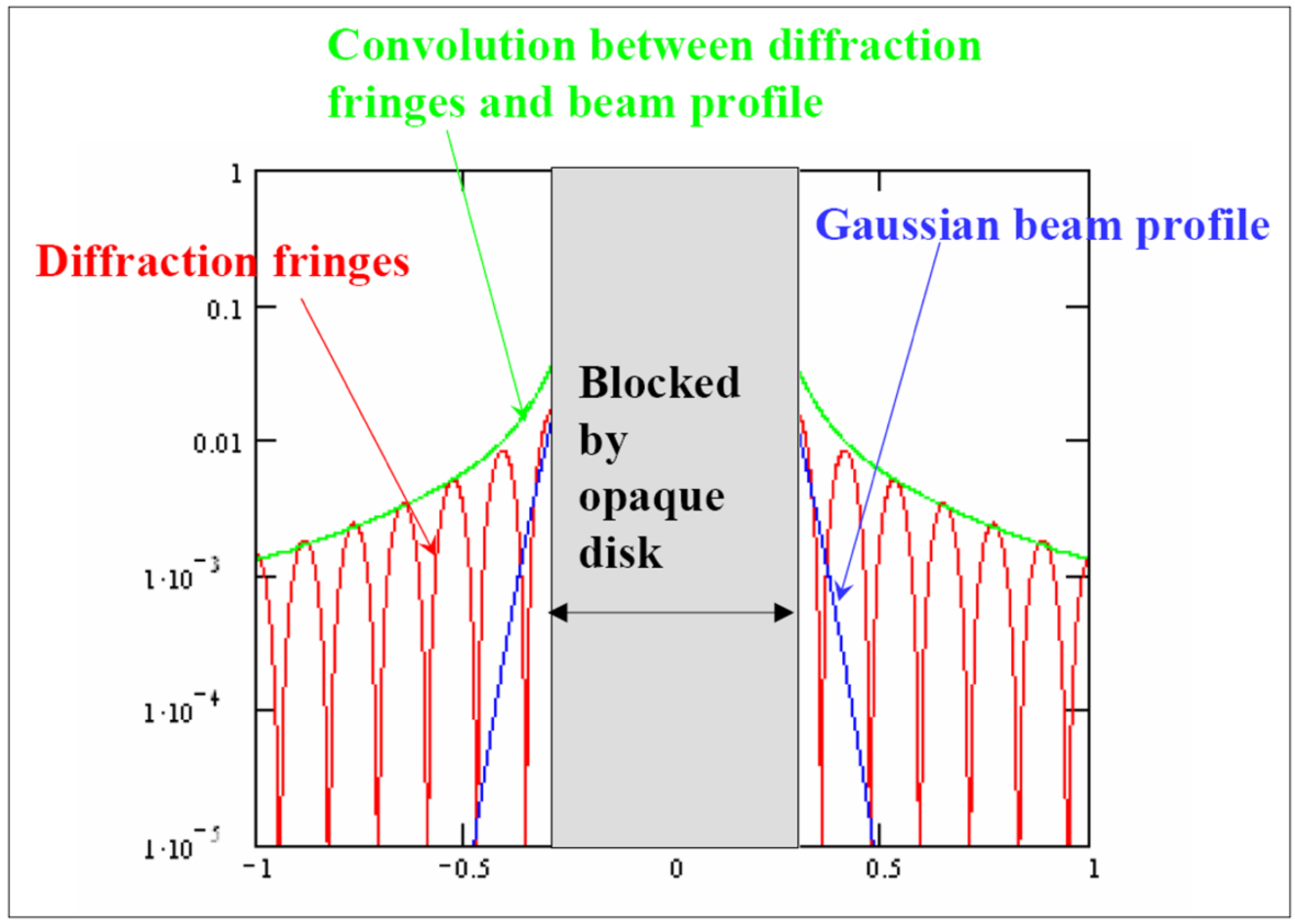

**Fig. 11:** Illustration of the effect of a mask in diffraction limited beam spot. The fringes over-shine beam halo of less than $10^{-3}$; from Ref. [70].

# 4 Longitudinal halo

The meaning of "longitudinal halo" can be divided into three different classes of different interests.

1) Neighbor Bunches or Bunch Purity: In time resolved experiments on synchrotron light sources one needs a clear signal from one bunch without any contributions from the adjacent (neighbor) buckets.
2) Beam in the Abort Gap: Superconducting hadron storage rings need a gap in the bunch train to have enough time for loading the dump kickers to ensure a clean beam dump. In case of a beam dump any particles in the gap will be lost around the ring risking a quench.
3) Coasting Beam: Experiments in colliders need very clear background conditions and precise time structures of the bunch crossings. Particles outside the main bunches may contribute to background as well as to undefined timing of the trigger counters in the experiment.

In particular the neighbor bunches have to be determined on a very low level of better than $10^{-6}$. Setups to measure the three different longitudinal halos are discussed in the following.

## 4.1 Bunch purity

Measurement of the sometimes special fill patterns of synchrotron light sources (rings) is important for time-resolved experiments. The adjacent buckets must not have any stored particles or, in reality, as few as possible. A method with very good time resolution (<< 1ns for a 500 MHz RF-System) and high dynamic range (more than six orders of magnitude) is necessary to sufficiently measure the contents of the neighbor buckets relative to the main bucket. The main mechanisms of filling neighbor buckets are [75]:

- *Quantum lifetime.* An electron is lost from a bucket by emitting a photon having a momentum larger than bucket height and can be re-captured in a neighboring bucket.
- *Lifetime determined by the vacuum pressure*. Electrons lose energy in collisions with residual gas molecules in the vacuum chamber and can be re-captured in a neighboring bucket.
- *Touschek effect.* Electrons in a bunch execute betatron oscillation with transverse momenta. When two electrons are scattered elastically, the transverse momenta can be transferred to the ongitudinal plane. The electrons can be re-captured by the forward or backward bucket, respectively.
- *Injection errors (energy, timing).* Incorrect timing and energy matching at injection can cause a growth on the both sides of the main buckets. This becomes even more important at topping up injections.

A typical setup to measure the bunch purity is a time-correlated single photon counting method (TCSPC).

### *4.1.1 Time-Correlated Single Photon Counting method (TCSPC)*

The arrival time of single photons emitted by electron bunches passing through a selected dipole in the storage ring can be measured with a resolution of < 1ns relative to a clock pulse which is synchronized to the bunch revolution frequency via the storage ring RF system. The photons are detected by a very fast Photomultiplier Tube (PMT) or an Avalanche Photo Diode (APD). Since the emitted photons have a too high flux for direct observation, x-ray photons scattered from a thin foil are used. Often a thin graphite foil of about 1mm is used which resists the thermal heat load induced by the direct photon beam (see Fig. 12). The detector must be carefully shielded against any stray light. The detector signals are amplified close to the detector by a fast amplifier. The amplified signal is analyzed using a Time-to-Digital-Converter (TDC) and a Multi-Channel-Analyzer (MCA). The time between the start signal coming from the detector and the stop signal from the next bunch clock or next revolution clock (depending on the required time scale and resolution) is measured and collected in a histogram (see Fig. 13). To reduce the influence of the so-called "walk" and to reduce the background due to electronic noise the amplified detector signal should be filtered by a constant-fraction-discriminator (CFD) [76].

A typical TDC-board used in an experiment at PETRA offers 4096 channels with minimum width below 40 ps and can work at maximum count rates up to 3 MHz (300 ns recovery time) [77]. To measure a histogram not affected by recovery-time and pile-up effects, the detector count rate should be limited to below 1.5% of the maximum rate. Assuming the PETRAII conditions with a bunch distance of 10 MHz, an adequate count rate of 10 kHz and a desired dynamic range of $10^7$; the time to resolve $1/10^7$ is about 100 s, with better statistic one needs 1000 s ≈ 16 min, which is quite a long time. A measurement of neighbor bunches resulting from a mismatched injection energy is shown in Fig. 14. Note the high dynamic range of about $10^7$ of this result. Other examples can be found in Refs. [78] and [79].

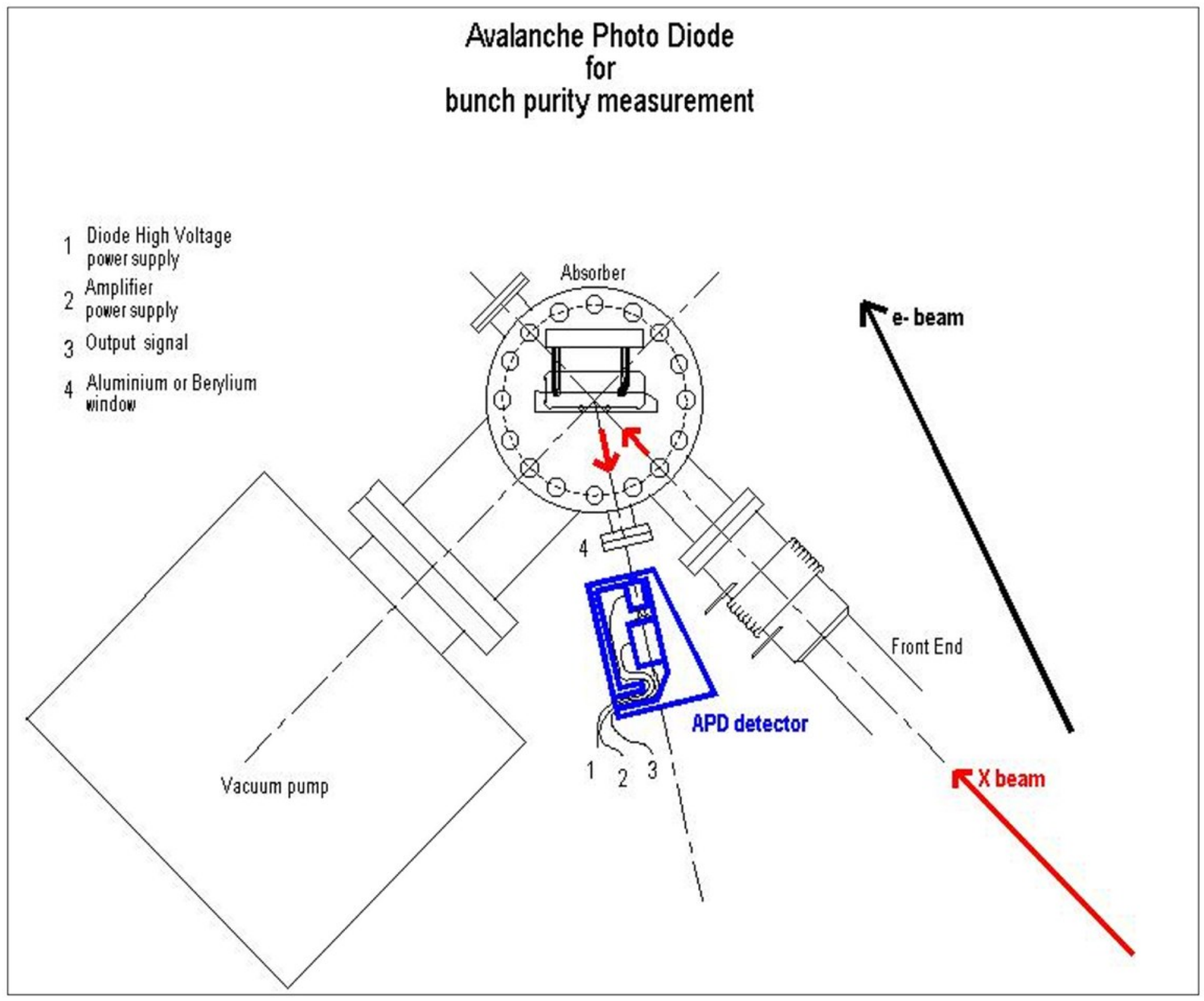

**Fig. 12:** A typical setup of the TCSPC method, here the ESRF Setup [87]

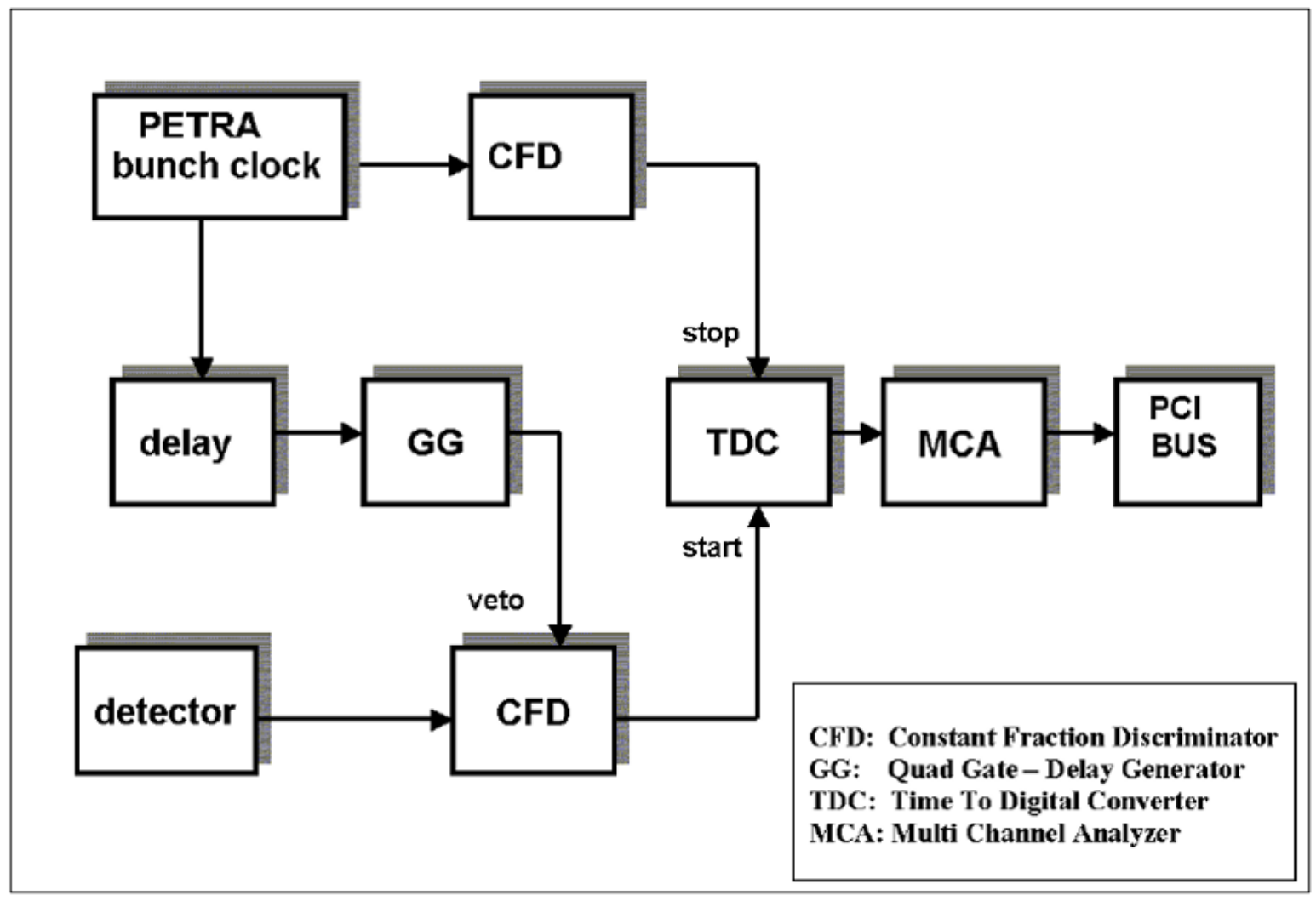

**Fig. 13:** Typical signal processing path for a TCSPC measurement [77].

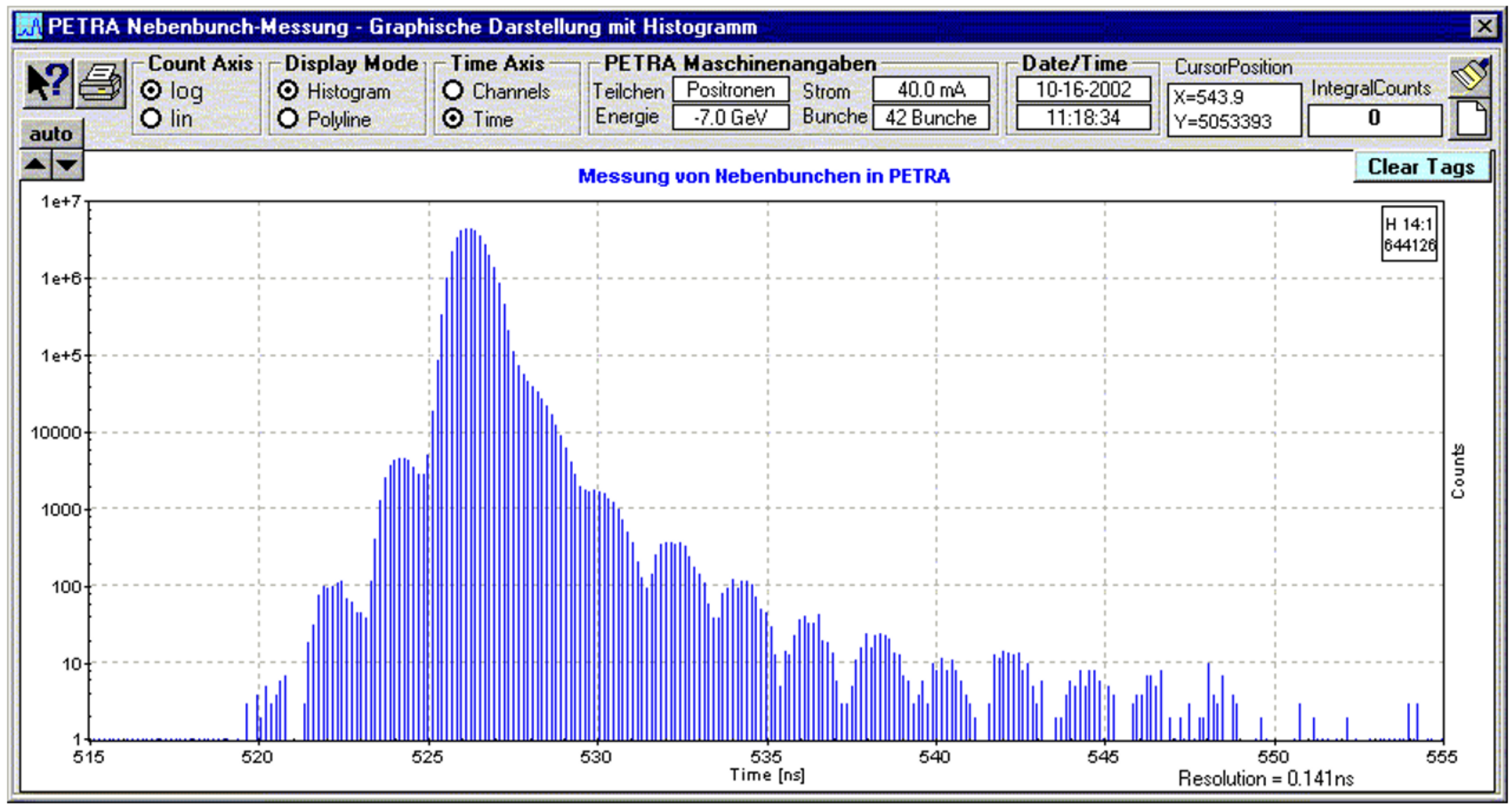

**Fig. 14:** Main bunch and parasitic bunches in PETRAII from 6ns before to some 20ns after the main bunch. A mismatch in energy led to many neighbor bunches [77].

*4.1.1.1 Improvements:*

1) Better TDC with higher resolution and less recovery time: A commercial TDC system has been tested successfully at Diamond [80]. An improved system from the same company (HydraHarp 400) is descript on the website of PicoQuant and used at HEPS [81]. This system improves the acquisition time with a time resolution down to 1 ps, a count rate of 12.5 MHz/channel and a dead time of <80ns. A common sync input for all channels permits the use of the system for TCSPC in forward start-stop mode at stable excitation sources up to 150 MHz.

2) Use of Micro Channel Plate Photo Multiplier Tube (MCP-PMT) for better detector timing resolution: A typical APD (used in Ref. [77]) with a sensitive area of 10 x 10 mm$^2$ , for example, the C30703F from Perkin Elmer (formerly EG&G) has a rise time of $\geq 1$ ns and a dark count rate of 20-500 counts per second. Note that smaller (x-ray) APDs can be faster with less dark count rates [82], [83]. A possible replacement for an APD can be e.g. a MCP-PMT Type R3809U-50 from Hamamatsu (used in [80], for example). It has a sensitive diameter of Ø 11 mm and a rise time of $t_{rise} = 0.15$ ns. This faster rise time allows a much better timing resolution for the measurement. Note that both devices have a typical dark count rate of about 100 c/s. This limits the dynamic range in a 100 ns interval to not larger than $10^7$. Note that recent developments in avalanche photomultiplier tube show a time resolution of 50 ps [84].

3) Since the analogue and digital electronic components became much faster, a digital sampling of individual bunches by a BPM even at high repetition rates became possible and can be used for very fast bunch filling pattern measurements [85].

4) Increase the dynamic range by an optical shutter: The dynamic rage of a simultaneous measurement of the main bucket and neighbor buckets is limited on the one side by the dark count rate of the detector and on the other by the dead time of the TDC. A reduction of the count rate of the main bunch by a known factor will extend the dynamic rage by orders of magnitude. Such a system was developed in [86] and is roughly described in the following (see also Fig. 15): A fast Pockels Cell is installed in the optical light path of the synchrotron radiation from a bending magnet. The timing of the Pockels Cell can be adjusted so that the light path is “open” for the neighbor bunches but “closed” for the main bunch. The measured extinction ratio between an “open” and “closed” Pockels Cell was measured to be about 200. By using two successive Pockels Cells the extinction ratio became $10^5$. That means that the main peak is suppressed by $10^{-5}$ due to the (limited) shutter efficiency. Therefore the dynamic range is extended by some orders of magnitude; a dynamic range of $\approx 10^{10}$ was reported in Ref. [86]. Also, the measuring time for the neighbor bunches at that level (!) improves only up to 500 s. The system is used continuously over days to measure the very small growth of bunch impurity during top-up operation at Spring-8.

*4.1.1.2 Instrumental effects:*

Note two instrumental limitations affecting the measured count rate in TCSPC experiments:

1) Pile-up: Assuming a gap in the bunch train and that the count rates are comparable with the dead time of the TDC, the probability for a photon from one of the first pulses to be detected is larger than that for the rest of the train [80] resulting in an uneven detection efficiency.

2) After-glowing: The R3809U-50 MCP-PMT showed some "after-glowing" [80]. Spurious pulses were counted at three different times after the main pulse (8, 45 and 100 ns) resulting in artificial neighbor bunches. These pulses limit the dynamic range in the most affected time-bins to only $10^5$.

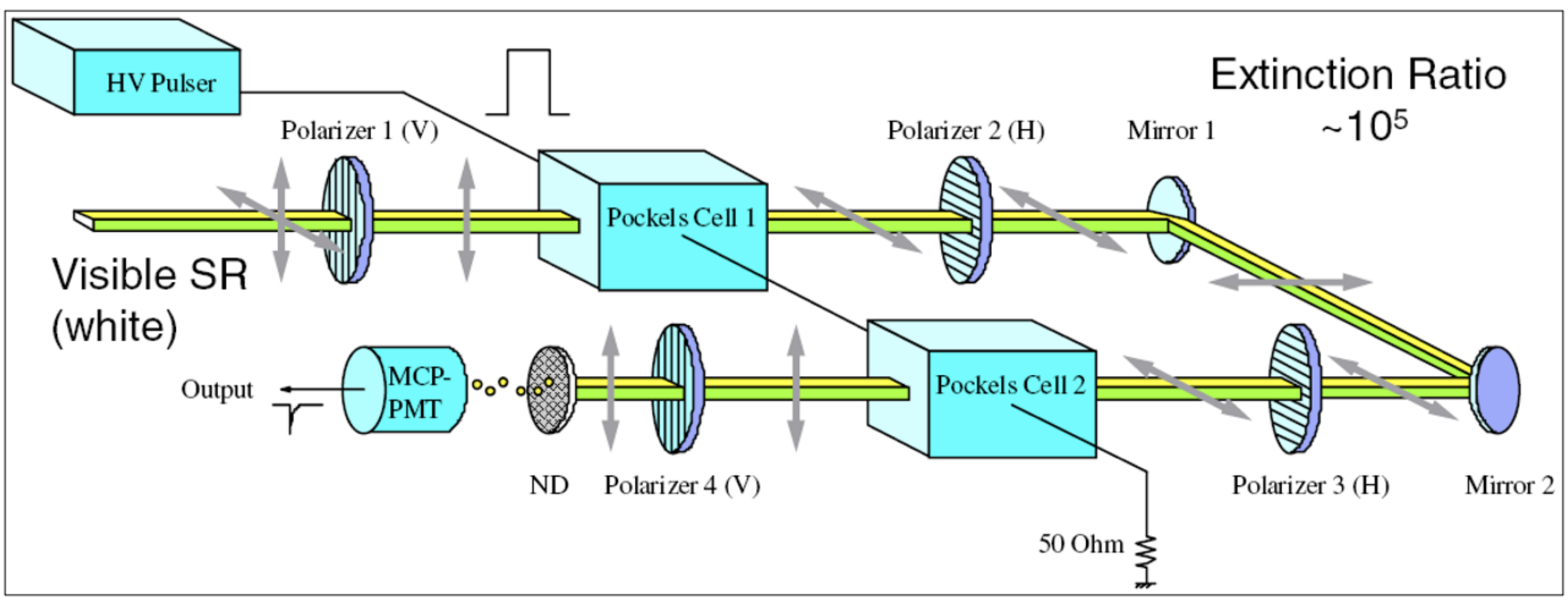

**Fig. 15:** Schematic drawing of the SPring-8 bunch purity monitors system [86].

## 4.2 Beam in the abort gap

At the end of a store of a superconducting hadron storage ring the beam has to be dumped in a safe way. The loading of the dump kicker to its design flat top value requires a certain time, typically a few μs. This time is known as the abort gap where no particles should be stored. Any beam in this gap (bunched or coasting beam) will spray around the machine if the dump kicker is fired causing some problems:

- Quenches (SC-magnets)
- Activation
- Spikes in experiments
- Equipment damage.

Reasons for beam in the abort gap can be

- Injection errors (timing)
- Debunching
- Diffusion
- RF noise/glitches
- Other technical problems.

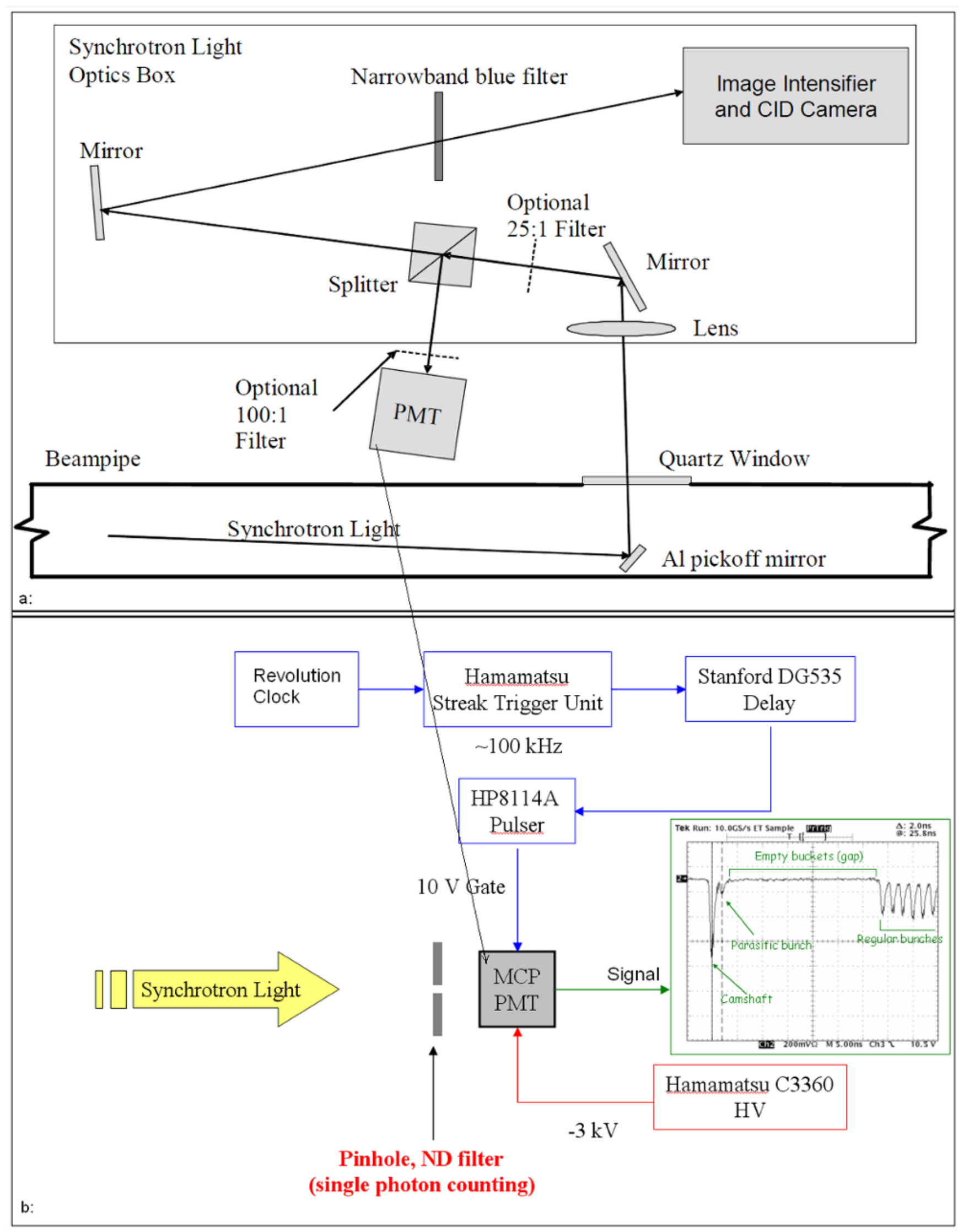

**Fig. 16:** Setup of the synchrotron radiation profile monitor at the Tevatron (a: upper part; [89]) and the block diagram of the beam in gap experimental setup (b: lower part, [92]). The revolution clock synchronize the trigger unit to the beam revolution frequency, the delay can move the gate across the abort gap. The pinhole and ND filter is added for TCSPC experiments.

Therefore, a continuous determination of the amount of beam in the gap is necessary to either clean the gap[1] or dump the whole beam before major problems arise. At high energy storage rings like the LHC, Tevatron, or HERA the presence of particles in the gap can be detected by means of the synchrotron radiation they emit, using the synchrotron radiation profile monitor port (Fig. 16). Note that in principle any other fast process, e.g. beam-induced gas scintillation or secondary electron emission or beam loss monitor signals can serve as a signal source, which are not limited to very high beam energy [88]. A fast and gate-able detector which is synchronized by the revolution frequency is most useful to avoid saturation due to the signal of the main bunches. Also, the detector has to be very sensitive in order to measure even a small amount of beam in the gap. Here the use of synchrotron radiation will be discussed as an example: An optical detector measures the time structure of the synchrotron radiation. A gated MCP-PMT is fast and sensitive enough to measure both components of the beam in the gap, the bunched (AC) and the unbunched (DC) components while an intensified gated CCD or CID camera integrates over many turns and measures the DC component only [89]. Typical gate rise times of about 1 ns are sufficient for this application. Often the display of the analog signal of the MCP-PMT versus the gate-time is sufficient but the dynamic range is limited to about $10^3$ due to the noise of the instrument. When using the gating technique, one has to take into consideration the maximum duty cycle of the instrument. A typical maximum duty cycle of 1% (e.g. Hamamatsu R5916U-50 MCP-PMT, see [90]) might not allow a complete gate over the whole gap at every turn. Therefore the gate repetition rate has to slow down or a shorter gate has to be moved across the gap [91]. In any case averaging over many turns helps to improve the sensitivity of the measurement. Some experiments are still ongoing to increase the dynamic range and the signal-to-noise ratio by applying the TCSPC Method (Section 4.1.1) to this measurement. First results with MCP-PMTs and APDs are reported in Refs. [92] and [93], more recent results in Refs. [94] and [95].

[1] Abort gap cleaning by, for example, fast kickers, resonant excitation, electron lens, etc. is not discussed in this lesson.

A completely different system to measure satellite bunches is the use of a fast wall current monitor. However, quite precise estimates measuring bunch populations below the $10^{-3}$ level require the compensation of the non-linear phase-delays, signal attenuation and recovery of the zero baseline, particularly if several bunches are circulating [96].

### 4.3 Coasting beam

Coasting beam is the part of the beam which is not captured by the RF system; their energy is not being replenished by the RF system. In electron storage rings electrons lose so much energy per turn that uncaptured particles are lost within very few turns. In high energy hadron storage rings uncaptured protons lose only a few eV per turn so that they can be stored for many minutes to hours. Uncaptured beam slowly spirals inward and is lost on the tightest aperture in the ring. RF noise, glitches or intra beam scattering can cause diffusion out of the RF-buckets leading to coasting beam. The total uncaptured beam intensity is a product of the rate at which particles leak out of the buckets and the time required for them to be lost. This kind of beam loss causes additional activation of the collimators as well as additional background in the experiments. Collider experiments in particular might suffer from this background; therefore, they are interested in measuring the amount of coasting beam. The accelerator specialists like to know the diffusion rate for a better understanding of the source of coasting beam [10], [12]. Of course, the methods discussed in Section 4.2 are also sufficient to determine the coasting beam, but more sensitive methods might be useful to measure small fractions of coasting beam in an appropriate time. Therefore, the experiments themselves can use their detectors and fast trigger equipment which have a very large detection efficiency as well as very small dead times. Detailed measurements of coasting beam are reported from HERA-B (HERA) and CDF (Tevatron) [97]-[99]. Both detectors use as the signal source the beam loss in the detector while HERA-B even increased the loss rate using its internal wire target. The time structure of the losses is measured by fast counters and TDCs (HERA-B; very similar to the TCSPC method discussed in Section 4.1.1, but with much more channels) or by integrating counts versus a sliding time interval (CDF). The time resolution of the HERA-B detector is good enough (≈ 0.1 ns) to measure also captured particles in neighbor buckets (Fig. 17).

But note that the signal source comes from the far transverse halo of the beam. Its time structure might differ strongly from the time structure of the beam core [97], especially because uncaptured beam slowly spiral inward. Therefore, a total determination of the amount of coasting beam will have a large uncertainty. As soon as the amount of coasting beam is large enough, an absolute determination can be done by comparing the AC and DC beam current monitors readings. The DC monitor measures all circulating particles while the AC monitor is sensitive only to the bunched beam component (Fig. 18). The calibration of both monitors to each other can be done just after finishing the acceleration where no coasting beam can survive.

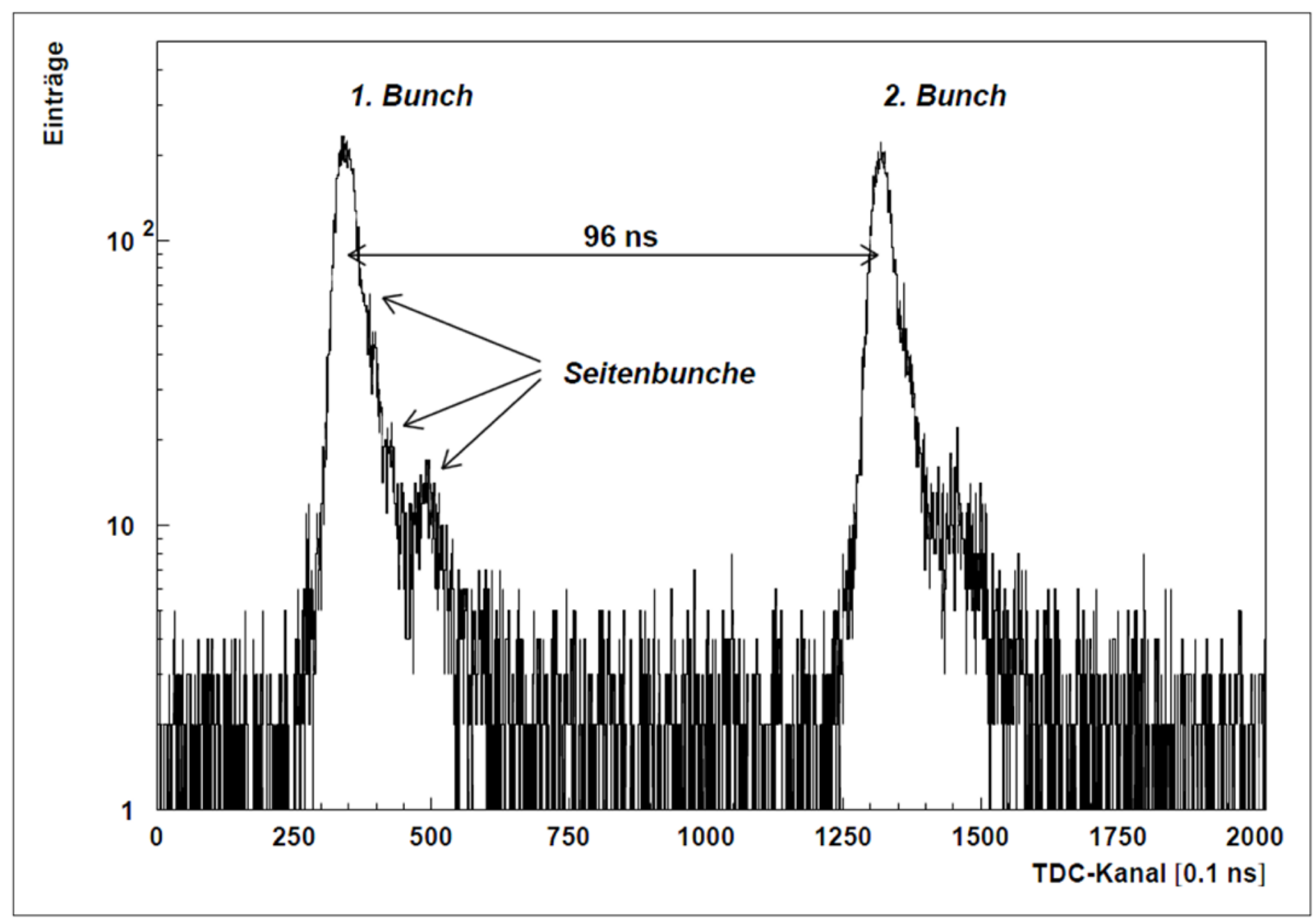

**Fig. 17:** Counts versus TDC channel in HERA-B [97]. Note the filling of the 500 MHz (2ns) neighbor buckets. The main bunches in HERA have a distance of 96 ns. The filled bins between the buckets indicate coasting beam.

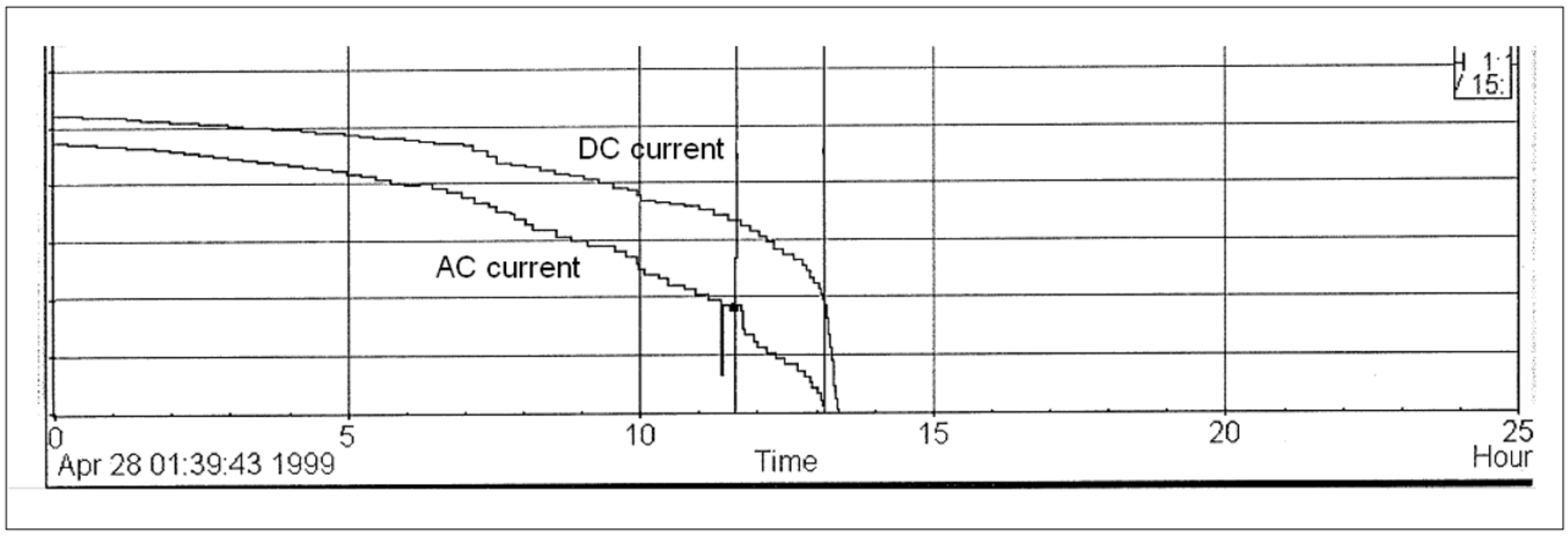

**Fig. 18**: DC beam current (includes coasting beam) and AC beam current (sum of all bunches) in HERA during an unusual store with a large amount of increasing coasting beam.

## 5 Summary

In this lesson transverse and longitudinal halo monitors have been discussed. Some useful definitions of halo have been given. The "state of the art" instrument for horizontal halo measurements is still the wire scanner. Its dynamic range is enhanced either by a counting readout scheme or by additional scrapers or detectors next to the wire. A dynamic range of better than $10^8$ has been achieved. Simple scraping methods might be more sensitive but suffer from the fact that they cannot determine the beam core. Optical methods using the synchrotron radiation readout by a CID camera or a coronagraph have the potential to reach even higher dynamic range.

Longitudinal halo: Bunch purity measurements with a dynamic range of more than $10^7$ were achieved with the Time-Correlated Single Photon Counting Method. The range can be extended to $10^{10}$ by introducing one or two Pockels cells into the light path which suppress the light from the main bunch by some orders of magnitude. For "beam in gap" measurements a fast electronic gate can be adequate.

At high energy proton accelerators the synchrotron radiation is mainly used as a signal source, together with a fast and gateable MCP-PMT or an APD. Often their analogue signal is sufficient but counting methods can improve the dynamic range. The measurement of coasting beam is often the task of the HEP experiments because of the huge sensitivity of their detectors. Their counting and triggering equipment is ideal to measure the time structure of beam losses in the detector. An absolute determination of the coasting beam has to be done by calibrated beam current monitors.

## Acknowledgements

Many thanks to my colleague G. Kube from DESY for adding useful comments to the manuscript. Many authors listed in the references provided me with their talks, pictures and data, which are all gratefully acknowledged.